\documentclass[twocolumn,amssymb, aps,nobibnotes,prd,showpacs]{revtex4-1}
\usepackage[colorlinks=true,citecolor=blue,urlcolor=blue,linkcolor=blue,hyperfigures=true]{hyperref}
\usepackage{graphicx}
\usepackage{bm,braket}
\usepackage{amsmath, amssymb}
\usepackage{color}

\newcommand{\inu}{{\imath\nu}}

\begin{document}

\title{Intersite electron correlations on inhomogeneous lattices:\\ a real-space dual fermion approach}
\author{Nayuta~Takemori$^{1,2}$, Akihisa~Koga$^3$, and Hartmut~Hafermann$^4$}%
\affiliation{
$^1$Center for Emergent Matter Science, RIKEN, Wako, Saitama 351-0198, Japan\\
$^2$Research Institute for Interdisciplinary Science, Okayama university, Okayama, 700-8530, Japan \\
$^3$Department of Physics, Tokyo Institute of Technology, Meguro-ku, Tokyo 152-8551, Japan\\
$^4$Mathematical and Algorithmic Sciences Lab, Paris Research Center, Huawei Technologies France SASU, 92100 Boulogne-Billancourt, France}
\date{\today}%

\begin{abstract}
We develop a real-space extension of the dual fermion approach.
This method is formulated in terms of real-space Green's functions and local vertex functions, which enables us to discuss local and nonlocal correlations in inhomogeneous systems with an arbitrary geometric structure.
We first demonstrate that the approach gives us reasonable results for a homogenous periodic system by taking into account onsite and nearest-neighbor intersite correlations. 
Moreover, we study the half-filled Hubbard model on the quasiperiodic Penrose lattice and clarify the role of intersite correlations for the Mott transition. The intersite correlations lead to a rich structure in local observables which is inherited from the quasiperiodic structure of the lattice.
\end{abstract}

\maketitle

\section{\label{sec:level1}Introduction}
Recently, particle correlations in inhomogeneous fermionic systems 
have drawn much interest. 
Typical examples are fermionic atoms 
in a harmonic potential~\cite{ultracold_gases} and
electron systems on ${\rm LaTiO_3/SrTiO_3}$ 
heterostructures~\cite{IZUMI200153,interface}.
In the former system, a transition from a metallic to a band insulating state occurs by introducing the trapping potential in its center part~\cite{inhomo_optical}, and in the latter system, metallic behavior appears in the heterostructure of the Mott insulator and the band insulator~\cite{ohtomo2002artificial}.
Moreover, quasiperiodic systems which do not exhibit translational but rotational symmetry ($\it e.g.$ 8- or 10-fold) 
have also been considered following the observation of quantum critical behavior 
in the quasicrystal Au$_{51}$Al$_{34}$Yb$_{15}$~\cite{Ishimasa,Deguchi12}.

Such inhomogeneous systems have theoretically been treated 
by the slave boson technique~\cite{PhysRevB.58.6692,PhysRevB.57.6937} and 
the real-space dynamical mean field theory 
(RDMFT)~\cite{Georges96,surface,cold1,cold2,cold3,Takemori,Takemura}, 
where interesting properties have been discussed such as an electronic reconstruction in a surface/interface system~\cite{surface}, a supersolid state in the optical lattice system with harmonic potential~\cite{cold3} and the local isomorphic distribution of local quantities~\cite{Takemori,Takemura}.
However, these methods can capture only local correlations. Antiferromagnetic and/or superconducting fluctuations enhanced by intersite correlations, which play an essential role for low temperature properties,
cannot be treated correctly.
Therefore, alternative theoretical techniques are necessary 
to discuss ground-state and low temperature properties in these systems.

Diagrammatic extensions of DMFT~\cite{Rohringer17,Kusunose} incorporate nonlocal correlations beyond the mean field level, while at the same time allowing a natural formulation in real space.
A corresponding generalization of the dynamical vertex approximation termed nano-D$\Gamma$A has been proposed in~\cite{Valli2010}. Calculations however were restricted to the DMFT approximation level~\cite{Valli2010,Valli2012}: the self-energy is assumed local albeit site-dependent, making the approach similar to RDMFT. A first application of nano-D$\Gamma$A including nonlocal correlations considered Hubbard nanorings. These nanostructures can be viewed as small one-dimensional systems with periodic boundary conditions and hence do not constitute a truly inhomogeneous system.

In this paper, we explain the real-space dual fermion (RDF) approach, which allows us to study intersite electron correlations in general inhomogeneous lattices beyond RDMFT. A first proof-of-principle has been given in~\cite{Takemoridualproc}.
Here we include, for the first time, nonlocal correlations via diagrammatic corrections to the RDMFT self-energy, while applying the approach to a truly inhomogeneous system.
We study the Mott transition in the Hubbard model on the square and the quasiperiodic Penrose lattices 
as examples of strongly correlated electron systems on homogeneous and inhomogeneous lattices. 
We then examine how local quantities are affected by the short-range electronic correlation effects at finite temperatures. We also observe the reconstruction of electronic profiles in the insulating region.

This paper is organized as follows. 
In Sec.~\ref{sec:rdf}, we introduce RDF.
In Sec.~\ref{sec:results}, we study the two-dimensional half-filled Hubbard model on the periodic square lattice and the Penrose lattice and show the efficiency of the RDF approach. The enhancement of the site-dependent renormalization is also addressed. A brief summary is given in the last section.

\section{Real-space dual fermion approach}
\label{sec:rdf}
The dual fermion approach is a diagrammatic extension of the
single site DMFT. The introduction of an auxiliary fermion field allows us to treat
intersite correlations beyond DMFT~\cite{Rubtsov,Otsuki,Hartmut}.
The method has successfully been applied to various models and lattices~\cite{Rohringer17}, leading, for example to a rich magnetic phase diagram on the triangular lattice~\cite{Li2014}, or a strong reduction of the critical interaction of the Mott transition compared to DMFT~\cite{Hafermann09}. The method even reproduces the non-mean field critical exponents~\cite{FK_critical_exponent,Hirschmeier2015}.
Introducing the auxiliary fermions in a real-space description enables us to consider nonlocal correlations in general inhomogeneous systems by solving an effective local problem for inequivalent sites.
In the following, we derive an action for the auxiliary field, explain resulting diagrams and give a detailed overview of RDF.

We start with the single-orbital Hubbard model for the correlated electron system on an arbitrary lattice.
The Hamiltonian is given by
\begin{align}
\label{H}
H = \sum_{\langle ij \rangle \sigma} t_{ij} c_{i\sigma}^{\dagger}c_{j\sigma} + \sum_{i}U_{i}{n_{i\uparrow}n_{i\downarrow}}-\sum_{i \sigma} \mu n_{i\sigma}.
\end{align}
where $\langle ij \rangle$ denotes the sum over pairs of sites, $c^{(\dagger)}_{i \sigma}$ is 
an annihilation (creation) operator of an electron at the $i$th site with spin $\sigma(= \uparrow , \downarrow)$ and $n_{i\sigma} = c^{\dagger}_{i\sigma} c_{i\sigma}$. Furthermore, 
$t_{ij}$ denotes the transfer integral between sites, $\mu$ the chemical potential and $U$ the Coulomb interaction. For simplicity, we take $t_{ij}$ equal to $-t$ when $i$ and $j$ are nearest neighbors and equal zero otherwise. In the calculations we further take the Coulomb interaction to be site-independent $U_{i}\equiv U$. Note that the approach is straightforwardly applied to a site-dependent interaction.
In the current approach it however must be local (that is, depend on a single site index), as will become apparent below. The discrete index $i$ runs over the total number of sites $N$.
The action for the single-orbital Hubbard model is given by
\begin{eqnarray}
\label{slat}
\displaystyle S[c^{*},c]=-\sum_{ij\nu\sigma} c^{*}_{i\nu\sigma}\{(\inu+\mu)\delta_{ij} -[\hat{t}]_{ij}\}c_{j\nu\sigma}, \nonumber \\
+ \sum_{i\nu}U_{i}n_{i\nu\uparrow}n_{i,-\nu\downarrow},
\end{eqnarray}
where $c^{(*)}_{i\nu \sigma}$ is a Grassmann number 
at the $i$th site with spin $\sigma(= \uparrow , \downarrow)$ and $\nu$ denotes the fermionic Matsubara frequency. In the following, we mark matrices in lattice indices by a caret. For example, $\hat{t}$ denotes the symmetric matrix of hopping amplitudes, and $[\hat{t}]_{ij}$ is its $i,j$-element.

An impurity model is introduced by formally adding and subtracting a local hybridization $[\hat{\Delta}_{\nu}]_{ij}=[\hat{\Delta}_{\nu}]_{ii}\delta_{ij}$ at each lattice site, which preserves the original action. 
This leads the original action to the following form
\begin{equation}
\label{slat_rew}
S[c^{*},c]=\sum_{i}S^{(i)}_{\rm{imp}}[c_{i}^{*},c_{i}] - \sum_{ij\nu\sigma}c^{*}_{i\nu\sigma}([\hat{\Delta}_{\nu}]_{ii}\delta_{ij}-[\hat{t}]_{ij})c_{j\nu\sigma},
\end{equation}
where the impurity action $S_{\rm{imp}}^{(i)}$ for site $i$ is given by
\begin{eqnarray}
\label{simp}
S_{\rm{imp}}^{(i)}[c^{*}_{i},c_{i}]=-\sum_{\nu\sigma} c^{*}_{i\nu\sigma}\{(\inu+\mu)-[\hat{\Delta}_{\nu}]_{ii}\}c_{i\nu\sigma} \nonumber \\
+ \sum_{\nu}U_{i}n_{i\omega\uparrow}n_{i,-\omega\downarrow}.
\end{eqnarray}

\begin{figure}[b]
\begin{center}
\includegraphics[width=19pc]{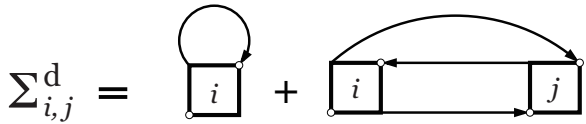}
\end{center}
\caption{\label{fig:dual_diagram}Diagrammatic representation of the first-order (left) and second-order (right) self-energy diagram of the dual fermion, Eq.~\eqref{eq:2nddiagram}.
}
\end{figure}

The auxiliary, so-called dual fermions ($f, f^{*}$) are introduced by applying the Hubbard-Stratonovich transformation to the second term on the right hand side of Eq.~\eqref{slat_rew}.
The partition function in terms of dual variables is rewritten by this transformation as below
\begin{eqnarray}
\label{rpartitionfunction}
Z= \int {\mathcal{D}}[f^{*},f] \int {\mathcal{D}}[c^{*},c] e^{-\sum_{i}S_{\rm{site}}^{(i)}[c_{i}^{*},c_{i};f_{i}^{*},f_{i}]} \nonumber \\
\times D_{f} e^{-\sum_{ij\nu\sigma}f^{*}_{i\nu\sigma}[\hat{g}_{\nu\sigma}^{-1}(\hat{\Delta}_{\nu\sigma}-\hat{t})^{-1}\hat{g}_{\nu\sigma}^{-1}]_{ij}f_{j\nu\sigma}},
\end{eqnarray}
where 
\begin{eqnarray}
\label{detf}
D_{f} &= \det[\hat{g}_{\nu\sigma}(\hat{\Delta}_{\nu\sigma}-\hat{t})\hat{g}_{\nu\sigma}].
\end{eqnarray}
Here, $\hat{g}_{\nu\sigma}$ is the diagonal matrix of impurity Green's functions with elements $g_{i\,\nu\sigma}\delta_{ij}$.
The local action $S_{\rm{site}}^{(i)}$ denotes the onsite part of the action, {\it i.e.}, the site index $i$ of the Grassmann number is unique for each term. 
$S_{\rm{site}}^{(i)}$ is further divided into two parts:
\begin{eqnarray}
\label{s_site}
&S_{\rm{site}}^{(i)}[c^{*}_{i},c_{i};f_{i}^{*},f_{i}]=S_{\rm{imp}}^{(i)}[c^{*}_{i},c_{i}] + S_{\rm{cf}}^{(i)}[c^{*}_{i},c_{i};f^{*}_{i},f_{i}],
\label{eq:Ssite}
\end{eqnarray}
where $S_{\rm{imp}}^{(i)}[c^{*}_{i},c_{i}] $ denotes the impurity action between physical fermions and $S_{\rm{cf}}^{(i)}[c^{*}_{i},c_{i};f^{*}_{i},f_{i}]$ represents the local coupling between the physical and the dual fermion given by
\begin{equation}
S_{\rm{cf}}^{(i)}[c^{*}_{i},c_{i};f^{*}_{i},f_{i}] =
\sum_{\nu\sigma} f^{*}_{i\nu\sigma}[\hat{g}_{\nu\sigma}^{-1}]_{ii}c_{i\nu\sigma}\!+\!c^{*}_{i\nu\sigma}[\hat{g}_{\nu\sigma}^{-1}]_{ii}f_{i\nu\sigma}.
\label{eq:Scf}
\end{equation}
By integrating out the physical fermion in $S_{\rm{site}}^{(i)}$ separately for each site, we obtain an effective action in terms of the dual fermion
\begin{equation}
S^{\rm{d}}[f^{*},f] = - \sum_{ij\nu\sigma}f^{*}_{i\nu\sigma} [(\hat{\mathcal{G}}_{\nu\sigma}^{\rm{d}})^{-1}]_{ij} f_{j\nu\sigma} + \sum_{i}V_{i}[f^{*},f].
\label{dualaction}
\end{equation}
Here we have defined the bare dual Green's function
\begin{eqnarray}
\hat{\mathcal{G}}^{\rm{d}}_{\nu\sigma}
=-\hat{g}_{\nu\sigma}[\hat{g}_{\nu\sigma}+(\hat{\Delta}_{\nu\sigma}-\hat{t})^{-1}]\hat{g}_{\nu\sigma}.
\label{eq:gdbare}
\end{eqnarray}
The local dual potential $V_{i}$ arises due to the particular form of Eq.~\eqref{eq:Ssite}: after expanding $\exp(S_{\rm{cf}}^{(i)}[c^{*}_{i},c_{i};f^{*}_{i},f_{i}])$, integrating out the fermions corresponds to an average over the impurity degrees of freedom because of the presence of $S_{\rm{imp}}^{(i)}[c^{*}_{i},c_{i}]$. This way, connected correlation functions of the impurity model are obtained which appear both in the bare dual Green's function~\eqref{eq:gdbare} as well as in the dual interaction $V$. As is common in the literature~\cite{Rohringer17}, we approximate $V$ by its leading term:
\begin{equation}
\label{eq:V}
V_{i}[f^{*}_{i},f_{i}] = - \frac{1}{4}\!\!\! \sum_{\nu\nu'\omega\, \sigma_{s}} \!\!\!\!\gamma^{\sigma_{1}\sigma_{2}\sigma_{3}\sigma_{4}}_{i\nu\nu'\omega}f_{i\nu+\omega,\sigma_{1}}^{*}f_{i\nu\sigma_{2}}f^{*}_{i\nu'\sigma_{3}}f_{i\nu'+\omega,\sigma_{4}}.
\end{equation}
Here, $\gamma^{\sigma_{1}\sigma_{2}\sigma_{3}\sigma_{4}}_{i\nu\nu'\omega}$ denotes a two-particle reducible vertex function depending on two fermionic ($\nu, \nu'$) and one bosonic frequency $\omega$ and defined by
\begin{align}
\label{gammadef}
\gamma_{i\,\nu\nu'\omega}^{\sigma\sigma\sigma'\sigma'} \!\!\!= \frac{g^{(4)\sigma\sigma\sigma'\sigma'}_{i\,\nu\nu'\omega}\!\!\!\! + \beta g_{i\,\nu+\omega\sigma}g_{i\,\nu\sigma}\delta_{\nu\nu'}\delta_{\sigma\sigma'} \!-\! \beta g_{i\,\nu\sigma}g_{i\,\nu'\sigma'}\delta_{\omega}
}{g_{i\,\nu+\omega,\sigma}g_{i\,\nu\sigma}g_{i\,\nu'\sigma'}g_{i\,\nu'+\omega\sigma'}},
\end{align}
where $\beta$ denotes inverse temperature.
Note that in the paramagnetic case that we consider, all non-zero spin configurations of the two-particle vertices and correlation functions can be obtained from the two components $\gamma^{\uparrow\uparrow\uparrow\uparrow}$ and $\gamma^{\uparrow\uparrow\downarrow\downarrow}$ in above equation.
The quantity $g^{(4)\sigma\sigma\sigma'\sigma'}_{i\,\nu\nu'\omega}$ is the local impurity two-body correlation function defined as
\begin{align}
\label{chi4def}
g_{i\,\nu\nu'\omega}^{(4)\sigma\sigma\sigma'\sigma'} &= \langle c_{i\,\nu+\omega,\sigma}c^{*}_{i\,\nu\sigma}c_{i\,\nu'\sigma'}c^{*}_{i\,\nu'+\omega,\sigma'}\rangle.
\end{align}
We consider the following local and nonlocal dual self-energy corrections generated by the auxiliary field potential term [Eq.~(\ref{eq:V})],
\begin{eqnarray}
[\hat{\Sigma}^{\rm {d}}_{\nu\sigma}]_{ij}=[\hat{\Sigma}^{\rm{d(1)} }_{\nu\sigma}]_{ii}\delta_{ij} + [\hat{\Sigma}^{\rm{d}(2)}_{\nu\sigma}]_{ij}.
\end{eqnarray}
Here, the resulting first-order self-energy diagram of the dual fermion (see Fig.~\ref{fig:dual_diagram}) is explicitly given by
\begin{eqnarray}
[\hat{\Sigma}^{\rm{d(1)}}_{\nu\sigma}]_{ii}= -T\sum_{\nu',\sigma'}
\gamma_{i\,\nu\nu'\omega=0}^{\sigma\sigma\sigma'\sigma'}[\hat{G}^{\rm{d}}_{\nu'\sigma'},]_{ii},
\label{eq:1stdiagram}
\end{eqnarray}
where $T$ denotes temperature and there is no summation over site indices $i$. The dual propagator of the dual fermion is defined as 
\begin{eqnarray}
\label{gdualdef}
[\hat{G}_{\nu\sigma}^{\rm{d}}]_{ij} &= -\langle{f_{i\nu\sigma}f_{j\nu\sigma}^{*}}\rangle.
\end{eqnarray}
Furthermore, the second order diagram reads
\begin{align}
\label{eq:2nddiagram}
&[\hat{\Sigma}^{\rm{d(2)}}_{\nu\sigma}]_{ij}=\nonumber\\
&-\frac{1}{2}T^{2}\!\!\!\!\sum_{\nu',\omega,\sigma'}\!\!\!\!
 \gamma_{i\,\nu\nu'\omega}^{\sigma\sigma\sigma'\sigma'}
[\hat{G}^{\rm{d}}_{\nu+\omega\sigma}]_{ji}
[\hat{G}^{\rm{d}}_{\nu'+\omega\sigma'}]_{ij}
[\hat{G}^{\rm{d}}_{\nu'\sigma'}]_{ji}
 \gamma_{j\,\nu'\nu\omega}^{\sigma'\sigma'\sigma\sigma}\nonumber\\
&-\frac{1}{2}T^{2}\sum_{\nu',\omega}
 \gamma_{i\,\nu\nu'\omega}^{\bar{\sigma}\sigma\sigma\bar{\sigma}}
[\hat{G}^{\rm{d}}_{\nu+\omega\bar{\sigma}}]_{ji}
[\hat{G}^{\rm{d}}_{\nu'+\omega\bar{\sigma}}]_{ij}
[\hat{G}^{\rm{d}}_{\nu'\sigma}]_{ji}
 \gamma_{j\,\nu'\nu\omega}^{\bar{\sigma}\sigma\sigma\bar{\sigma}}.\nonumber\\
\end{align}
Also, there is no summation over site indices. In the following, we drop spin indices where possible since we consider the paramagnetic case.
The dual Green's function is updated via the self-consistent equation of the dual fermion propagator
\begin{equation}
(\hat{G}^{d}_{\nu})^{-1}=(\hat{\mathcal{G}}^{d}_{\nu})^{-1}-\hat{\Sigma}^d_{\nu},
\end{equation}
until the dual self-energy is converged.
After the convergence of the dual Green's function and self-energy, the physical self-energy of the system is obtained from
\begin{align}
[\hat{\Sigma}
_{\nu}]_{ij} &=[\hat{\Sigma}_{\rm{imp},\nu}]_{ii}\delta_{ij}
+ [(\hat{1} + \hat{\Sigma}^{\rm{d}}_{\nu}\hat{g}_{\nu} )^{-1} \hat{\Sigma}^{\rm{d}}_{\nu}]_{ij}.
\label{eq:sigma}
\end{align}
\begin{figure}[htb]
\begin{center}
\includegraphics[width=\linewidth]{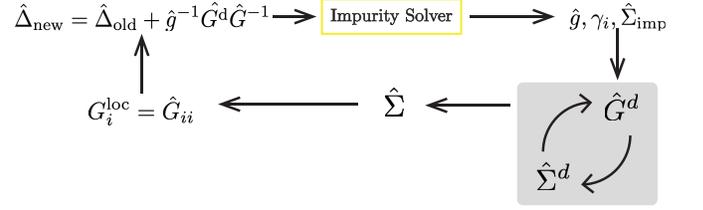}
\caption{\label{fig:dual_fermion}(Color online) The self-consistent loop of RDF. 
}
\end{center}
\end{figure}
The system Green's function $\hat{G}_{\nu}$ is constructed similarly as in RDMFT according to
\begin{equation}
\label{eq:glatinv}
\hat{G}_{\nu}=[\hat{\cal G}_{0 \nu}^{-1}-\hat{\Sigma}_{\nu}]^{-1},
\end{equation}
where ${\cal G}$ denotes the bare Green's function.
If we neglect dual self-energy corrections, Eq.~\eqref{eq:sigma} reduces to $[\hat{\Sigma}_{\nu}]_{ij} =[\hat{\Sigma}_{\rm{imp},\nu}]_{ii}\delta_{ij}$, and above equation indeed reduces to conventional RDMFT if the hybridization function is chosen accordingly. Therefore, the hybridization function must be updated as follows (we drop the frequency index for simplicity),
\begin{eqnarray}
[\hat{\Delta}_{\rm{new}}]_{ii} &= [\hat{\Delta}_{\rm{old}}]_{ii} + [\hat{g}^{-1}]_{ii} - [\hat{G}^{-1}]_{ii},
\end{eqnarray}
which implies that upon self-consistency, $\hat{g}_{ii}=\hat{G}_{ii}$.
A suitable self-consistency condition for RDF calculations is obtained by rewriting this equation in terms of the bare dual Green's function~\eqref{eq:gdbare} as follows,
\begin{eqnarray}
[\hat{\Delta}_{\rm{new}}]_{ii} &= [\hat{\Delta}_{\rm{old}}]_{ii} + [\hat{g}^{-1}]_{ii} [\hat{\mathcal{G}^{\rm{d}}}]_{ii} [\hat{G}^{-1}]_{ii}.
\label{eq:sc}
\end{eqnarray}
This implies that RDMFT corresponds to $[\hat{\mathcal{G}^{\rm{d}}}]_{ii} = 0$, or equivalently, setting the leading order diagram~\eqref{eq:1stdiagram} to zero without considering higher-order diagrams. RDMFT corresponds to an ensemble of noninteracting dual fermions.
When higher order corrections to the dual self-energy are taken into account, we still require the leading order diagram to be zero, which straightforwardly leads to the self-consistency condition of the RDF approach,
\begin{eqnarray}
\label{eq:self-consistent}
[\hat{G^{\rm{d}}}]_{ii} = 0.
\end{eqnarray} 
The self-consistent loop is continued iteratively until Eq.~\eqref{eq:self-consistent} is satisfied. This can be achieved using~\eqref{eq:sc} with $\hat{\mathcal{G}^{\rm{d}}}$ replaced by $\hat{G^{\rm{d}}}$.

We note that the above formulation of the RDF is equivalent to the dual fermion approach formulated in momentum-space~\cite{Rubtsov} when local quantities are site-independent, self-energies at all length scales are taken into account and periodic boundary conditions are enforced. This can be used as a simple test case.
In this paper, we only consider short-range correlations between nearest neighbor~(NN) sites which are supposed to be dominant:
\begin{equation}
[\hat{\Sigma}^{\rm{d}(2)}]_{ij} \sim [\hat{\Sigma}^{\rm {d}(2)}]_{ij}\delta_{\langle ij \rangle}.
\end{equation}
In the following, we use the transfer integral $t$ as the unit of energy, and we focus on the paramagnetic solution with $\langle n_{i \sigma} \rangle=0.5$ independent of $i$ to focus on the Mott transition at low temperature as an example of strong electron correlation effects.
Accounting for symmetries in case of the square lattice, the impurity models are solved  for inequivalent sites using a numerically exact hybridization expansion continuous-time quantum Monte Carlo algorithm~\cite{Hafermann2013} (CTHYB). In the case of the Penrose lattice, all sites are a priori inequivalent except for the rotational symmetry ($C_{5v}$). In that case we solve 13 inequivalent sites in the Penrose Hubbard model on a large cluster of 86 sites, so that boundary effects are expected to be small.

\section{Results}
\label{sec:results}
\subsection{Short-range correlation effects in the two-dimensional Hubbard model}
\label{sec:Square}
\begin{figure}[htb]
\includegraphics[width=\linewidth]{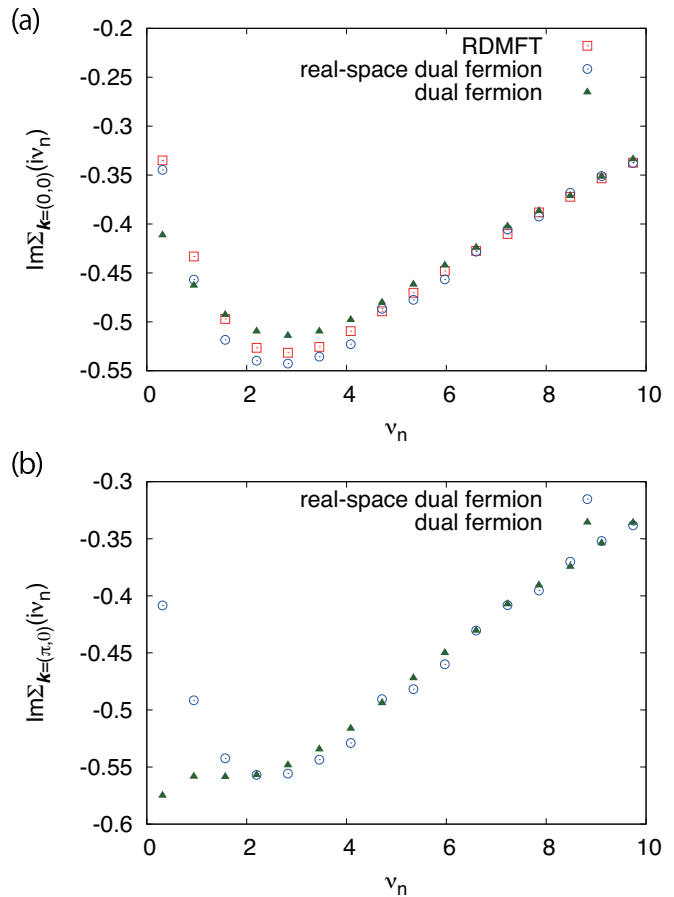}
\caption{\label{fig:SEsquare}(Color Online) The imaginary part of the system self-energy $\Sigma_{\bm{k}=(0,0)}(\inu_n)$ obtained by RDMFT, the RDF and dual fermion approach formulated in momentum-space (a)
and $\Sigma_{\bm{k}=(\pi,0)}(\inu_n)$ obtained by the real-space dual fermion approach and dual fermion approach formulated in momentum-space (b) for $U=4$ at $T=0.1$ on the square lattice with 16 sites.
The unit of energy is set to be $t$.
}
\end{figure}

Firstly, we study the two-dimensional square lattice Hubbard model with 16 sites under periodic conditions and demonstrate how the RDF captures short-range correlation effects. 
Figure~\ref{fig:SEsquare}(a) shows the system self-energy $\Sigma_{\bm{k}}(\inu_n)$  for ${\bm k}=(k_x,k_y)=(0,0)$ at $U=4$ by using RDMFT, RDF and the conventional dual fermion approach formulated in momentum-space (DF).
We find that these results are in good agreement with each other in the intermediate energy region ($\nu_n>6$).
On the other hand, they deviate from each other in the low energy region ($\nu_n<1$).
We find that our results are located between those obtained by RDMFT and DF formulated in momentum-space. This is because our RDF method does not take into account all intersite correlations (which lower the renormalization factor), but only some ({\it i.e.}, those between NN sites) beyond RDMFT.
Similarly, the crossing observed in the system self-energy $\Sigma_{\bm{k}}(\inu_n)$  for ${\bm k}=(k_x,k_y)=(0,0)$ at $n\sim 3$ is caused by restricting the second-order diagram to NN sites which leads to an underestimation of longer-range correlations.
We note that the off-diagonal component of the system self-energy for ${\bm k}=(k_x,k_y)=(\pi,0)$ displays  the same tendency as shown in Fig.~\ref{fig:SEsquare}(b) whereas that obtained by RDMFT is equal to zero.

\begin{figure}[htb]
\includegraphics[width=\linewidth]{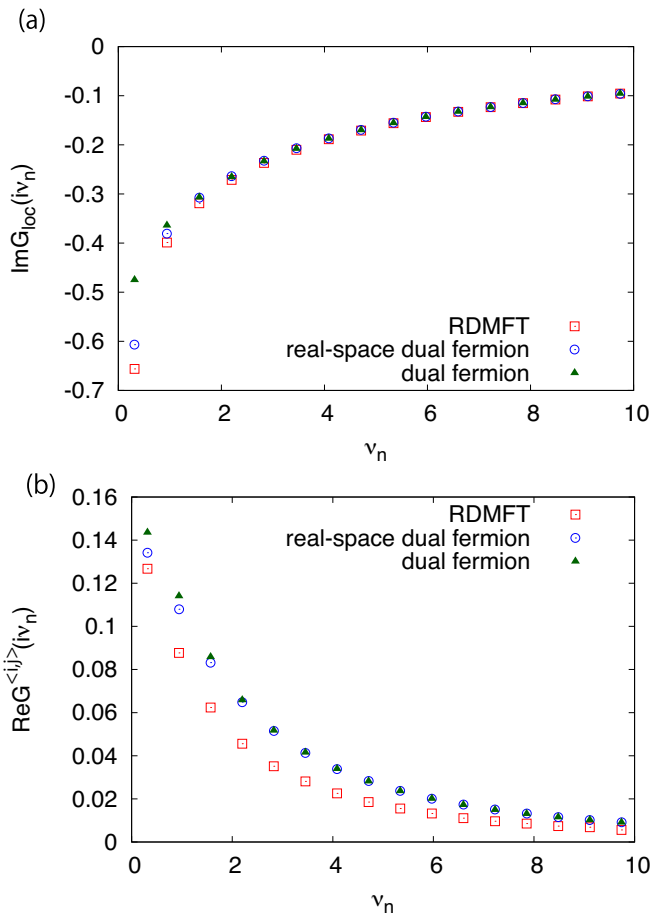}
\caption{\label{fig:gloc}(Color Online) The imaginary part of the local Green's function $G_{\rm loc}(i \nu_n)$ (a) and the real part of the off-site Green's function between nearest-neighbor sites $G^{\langle i,j \rangle}(\inu_n)$ (b) obtained by RDMFT, the real-space dual fermion approach and dual fermion approach formulated in momentum-space for $U=4$ at $T=0.1$ on the square lattice with 16 sites.
For the dual fermion approach formulated in momentum-space, we consider intersite correlations between all sites, whereas we consider short-range correlations only between NN sites in the real-space dual fermion approach.
The unit of energy is set to be $t$.
}
\end{figure}

We also calculate the system Green's function at $U=4$ as shown in Fig.~\ref{fig:gloc}.
For the local part [Fig.~\ref{fig:gloc}(a)], we find that our results are located between those obtained by RDMFT and the dual fermion approach formulated in momentum-space. 
We note that results obtained by RDF coincide numerically with that obtained by dual fermion approach formulated in momentum-space when we restrict second-order self-energy diagrams only between NN sites for both methods (not shown).
Results obtained by our RDF method are expected to approach that obtained by the dual fermion approach formulated in momentum-space when considering longer-range correlations. 
For the off-site component [Fig.~\ref{fig:gloc}(b)] between NN sites, the results show that short-range correlations are correctly taken into account. Although a small difference between results obtained by the RDF approach and DF appears in the low energy region ($\nu_n \leq 3$), results show good agreement in the higher energy region ($\nu_n  >3$), which clearly deviate from the results obtained by RDMFT.

\begin{figure}[t]
\includegraphics[width=\linewidth]{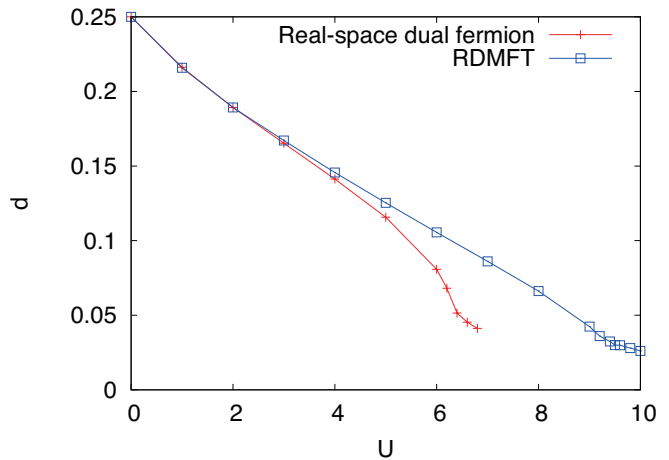}
\caption{\label{fig:SquareUvsd} 
(Color Online) Double occupancies in the Hubbard model on the square lattice with a periodic boundary condition obtained by means of RDF and RDMFT when $T=0.1$.
The unit of energy is set to be $t$.}
\end{figure}

To see how the difference in the system self-energy and the system Green's function affects the nature of Mott metal-insulator transition in the system, we show the interaction dependence of double occupancies $d=\langle n_{ \uparrow} n_{\downarrow} \rangle$ in Fig. \ref{fig:SquareUvsd}.
In the noninteracting case $U=0$, the normal metallic state is realized
with $d=0.25$.
By introducing the interaction, the double occupancy obtained by both methods 
monotonically decreases, similarly as in the conventional Hubbard model~\cite{single}.
The system self-energy obtained by the RDF approach results in stronger suppression of the mobility of electrons compared to RDMFT~\cite{Takemoriproc}, as clearly shown in the double occupancy obtained by in RDF at $U\geq3$.
At last, the first-order phase transition to the Mott  insulator occurs at $U_{c2}^{\rm periodic}$.
The transition point is deduced as $U_{\rm cross}^{\rm periodic}\sim9.4$ (RDMFT) and $U_{\rm c2}^{\rm periodic} \sim 6.4$ (RDF) at $T=0.1$.
The transition point $U_{c2}^{\rm periodic}\sim 6.5$ has been obtained in conventional DF~\cite{Hartmut}, $U_{c2}^{\rm periodic}=6.5$ in DCA~\cite{Werner_private}, the crossover $U_{\rm cross}^{\rm periodic}\sim 6$ by CDMFT~\cite{square_CDMFT} and $U_{\rm cross}^{\rm periodic}\sim 7$ by quantum Monte Carlo simulation~\cite{squareQMC} at the same temperature.
We conclude that RDF properly captures intersite correlations into account. While the corrections are smaller in magnitude compared to the conventional dual fermion approach in terms of local quantities, it gives us an improved picture of the Mott physics beyond RDMFT. In particular, the Mott transition occurs at a similar value of the critical interaction in RDF and DF.
The short-range correlation leads to a large value of the effective mass so as to form the Mott insulator at lower interaction strength~\cite{square_CDMFT}. 

In this section, we have investigated the half-filled Hubbard model on the square lattice under a periodic boundary condition and showed the efficiency of RDF by calculating local and off-site quantities. Furthermore, the obtained Mott transition point is consistent with findings obtained by other (cluster) extensions of DMFT and a numerically exact method.
These facts lead us to consider short-range correlation effects in inhomogeneous systems, which is addressed in the next section. 

\begin{figure}[t]
\includegraphics[width=\linewidth]{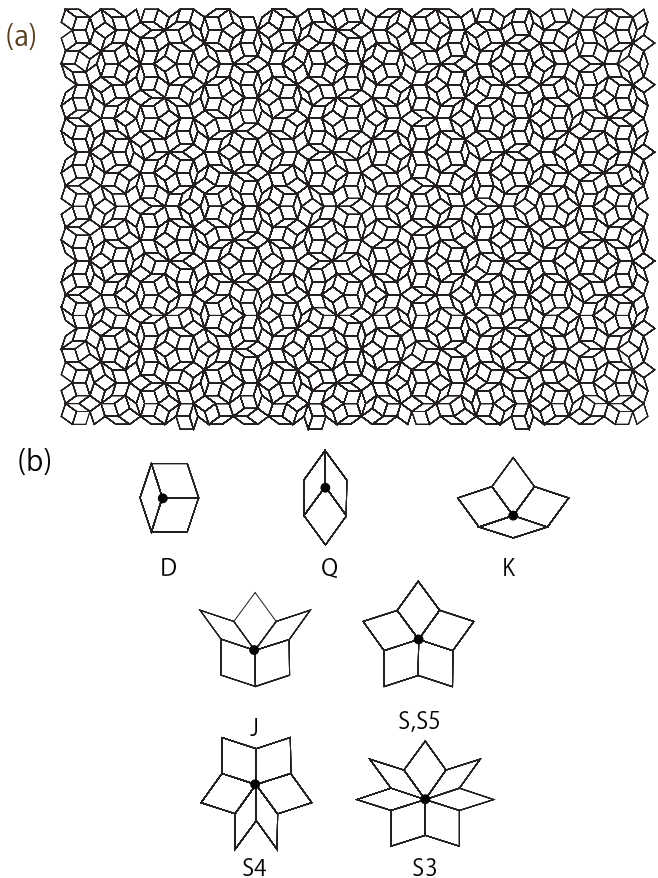}
\caption{\label{fig:Penrose_lattice} 
(Color Online)(a) Penrose lattice (b) and possible vertex patterns with coordination numbers 3 (D,Q), 4 (K), 5 (J, S, S5), 6 (S4) and 7 (S3) in the Penrose lattice.}
\end{figure}

\subsection{Short-range correlation effects in the Hubbard model on the Penrose lattice}
\label{sec:Penrose}
\begin{figure}[htb]
\includegraphics[width=\linewidth]{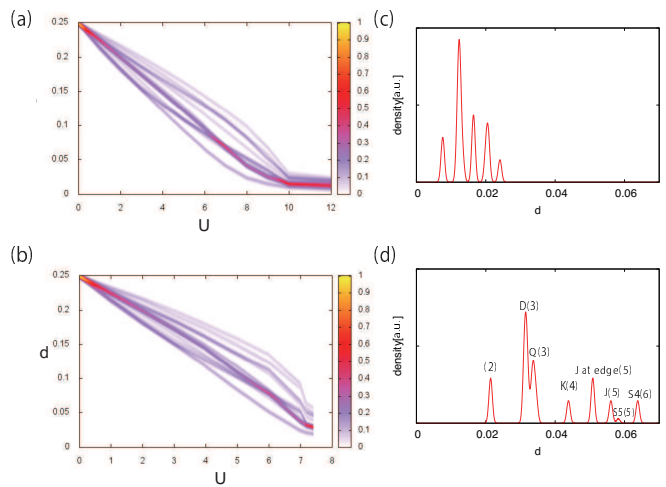}
\caption{\label{fig:Penrose}(Color online) Density plot of double occupancies obtained by RDMFT (a), and the real-space dual fermion approach (b) for 86 sites in the Penrose lattice
as a function of the interaction when $T=0.1$. (c) and (d) show their cross sections in the insulating regime [at $U=10.0$ for RDMFT (c) and $U=7.2$ for RDF (d)]. Labelled letter in (d) denotes corresponding vertex pattern and coordination number of each peak.
The unit of energy is set to be $t$.
}
\end{figure}

Next, we study the half-filled Hubbard model on the Penrose lattice~\cite{Takemori} with 86 sites
and show how short-range correlations affect the electronic properties in an inhomogeneous system. 
The Penrose lattice that we treat here is a vertex model where a site is placed on each vertex of the rhombuses as shown in Fig.~\ref{fig:Penrose_lattice}(a), and thus bipartite. The coordination number in the lattice ranges from 3 to 7 and each vertex pattern can be divided into eight classes (Fig.~\ref{fig:Penrose_lattice}(b)). We note that the cluster we employ only contains the lattice sites with its coordination number ranging from 3 to 6 and some sites with two connecting bonds appear only at the edge of the system. This diversity of the effective hopping enables interesting phenomena such as an unconventional weak-coupled superconductivity~\cite{Sakai} and an exotic magnetic pattern~\cite{koga-tsunetsugu}.
Figure~\ref{fig:Penrose} shows the distribution of the site-dependent double occupancy $d_i=\langle n_{i \uparrow}n_{i \downarrow} \rangle$ in the system with 86 sites obtained by RDMFT and RDF at the temperature $T = 0.1$.
In the metallic region ($U<4$), it is found that the quantities range within a certain small width since short-range correlation effects are not important in this region. 
In RMDFT, a crossover from a metallic to an insulating state occurs at $U_{\rm cross} \sim 10$.
On the other hand, the Mott transition occurs simultaneously for all sites at $U_{c}=7.2$ in RDF, which is significantly lower than the interaction of the crossover obtained by RDMFT. This tendency is similar to the case in the homogeneous system, which indicates that the RDF approach gives reasonable results in inhomogeneous systems.

We further observe that intersite correlation effects are important even in the regime beyond the Mott transition point $(U>U_{c})$. The distribution of the double occupancy obtained by means of RDMFT is grouped into five classes as shown in Fig.~\ref{fig:Penrose}(c). These peaks are characterized by the coordination number of the sites which ranges from 2 to 6 from left to right. Clearly, in RDMFT the position of a peak depends solely on the coordination number, while its height corresponds to the frequency of occurrence of the patterns with the respective coordination number.
This corresponds to the fact that local quantities are classified by their coordination number at each site in the atomic limit~\cite{Takemori}.

On the other hand, in the case of RDF, the distribution clearly has more structure and displays a certain width depending on coordination number and vertex pattern as shown in Fig.~\ref{fig:Penrose}(d).
We find that the distributions for the sites with coordination numbers 3 and 5 are grouped into two classes depending on two vertex patterns in terms of Bruijn's notation $\{D,Q\}$ and $\{J , S, S5\}$~\cite{Bruijn1,Bruijn2} as shown in Fig.~\ref{fig:Penrose_lattice}(b). 
Owing to the fact that the dual fermion self-energy consists of vertex functions localized at the impurity site and its NN site, local quantities not only depend on the coordination number of the respective site, but also the coordination number of its NN sites.
These findings indicate that the local geometry of the vertices affects electronic profiles dramatically in the  insulating regime and at intermediate coupling. 

\begin{figure}[t]
\includegraphics[width=\linewidth]{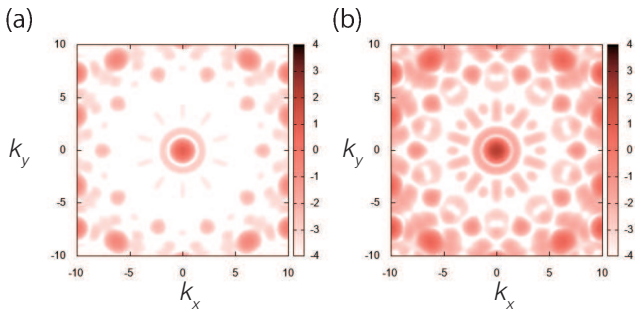}
\caption{\label{fig:diff}(Color online) Electron profiles of $\log_{10} |d_{\bm k}|^2$ for 86 sites in the Penrose lattice in the insulating regime obtained by RDMFT (a), and the real-space dual fermion (RDF) approach (b) when $T=0.1$.
}
\end{figure}

To see how the geometry of the lattice affects the distribution of local quantities obtained by RDMFT and RDF, we show electron profiles of double occupancies in the insulating regime [$U=10.0$ for RDMFT (a) and $U=7.2$ for RDF (b)] when $T=0.1$ as shown in Figure~\ref{fig:diff}.
To clarify how the site-dependent distribution of local quantities appears in the diffraction pattern,
we calculate the amplitude of $d_{\bf k}$, 
where
\begin{eqnarray}
d_{\bf k}&=&\frac{1}{N}\sum_i d_ie^{i {\bf k}\cdot {\bf r}_i}.
\end{eqnarray}
Again we find a much richer structure and additional peaks in $|d_{\bf k}|^2$ in the case of RDF.
Since the local quantities beyond the Mott transition point only depend on the coordination number of their respective sites in the case of RDMFT~\cite{Takemori}, this clearly indicates that the distribution of local quantities is characterized by the site geometry beyond the nearest neighbors in RDF.

In this section, we have investigated the half-filled Hubbard model on the Penrose lattice.
We have discussed the Mott transition and showed that RDF successfully captures short-range correlations when we consider self-energy diagrams between NN sites.
Moreover, we have shown that the short-range correlation effects lead to a much richer structure in the distribution of double occupancies and the electron density profile in RDF compared to RDMFT. This is qualitatively different compared to the case when we consider only local electron correlations in a quasiperiodic system and suggests that unconventional physical behavior characteristic of the quasiperiodic geometry may appear.
Further considerations of longer-range correlations in a quasiperiodic system is a possible application, which, for example, is indispensable for investigations of the origin of the quantum critical behavior of the quasicrystal Au$_{51}$Al$_{34}$Yb$_{15}$~\cite{Deguchi12}.

\section{Conclusions}

We have developed the real-space dual fermion approach, which for the first time has allowed us to investigate intersite electron correlation effects in strongly correlated inhomogeneous systems.
Nonlocal correlations between sites are included through diagrams in terms of auxiliary field fermions, leading to a diagrammatic extension of real-space DMFT.

We have studied the half-filled Hubbard model on the square lattice with a periodic boundary condition and discussed the Mott transition in this system.
Thanks to the real-space formulation, diagrams up to certain length scales can be selectively included to study their effects. By restricting the self-energy to nearest-neighbor terms, we found that significant corrections to the local Green's function stem from longer-ranged corrections, while the strong reduction of the critical interaction of the Mott transition compared to DMFT is already determined by short-range (nearest-neighbor) dynamical correlations. The Mott transition point is close to that obtained by means of other numerical techniques which include intersite correlation effects. This indicates that the real-space dual fermion approach gives a phase diagram which is consistent with the previous findings. 

A similar reduction of the critical interaction appears in the half-filled Hubbard model on the Penrose lattice. In addition we found that the inclusion of intersite correlations in an inhomogenous system leads to a much richer structure in the distribution of local quantities such as the double occupancy. This is different from RDMFT, where such quantities depend on the coordination number. By including nearest-neighbor correlations, local quantities also depend on the environment of nearest neighbors. This suggests that longer-range correlations yield an even larger site-dependent renormalization and richer structure in the Penrose lattice, which should affect physical behavior characteristic of the quasiperiodic system at very low temperatures.

Including self-energy corrections at larger or even all length scales is conceptually simple, but technically difficult because it requires summation over a large number of internal degrees of freedom in the diagrams, or the inversion of large matrices. We therefore leave this task for future work. The computational effort may be reduced using the so-called intermediate representation of observables~\cite{shinaoka2017compression}. It would be interesting to see whether longer-ranged corrections would lead inequivalent sites to undergo a Mott transition at different parameters.
 
We conclude that the real-space dual fermion approach captures into account intersite correlations in inhomogeneous systems, which leads to a rich structure of electronic properties resulting from the geometric shape of the underlying system and is expected to give rise to new phenomenon.
This method allows us to study intersite correlation effects in various other inhomogeneous systems such as cold atoms in a trapping potential, nanoscopic systems, interfaces and surfaces, impurities, molecules, topological insulators and quasiperiodic lattices.

\begin{acknowledgments}
The authors would like to thank A. I. Lichtenstein, J. Otsuki and P. Werner for valuable discussions. 
This work was partly supported by JSPS KAKENHI Grant No. 15J12110 and 16H07447 (N. T.) and 16H01066 and 17K05536 (A. K.).
Part of the computations was carried out on TSUBAME2.0
at Global Scientific Information and the Computing Center
of Tokyo Institute of Technology and at the Supercomputer
Center at the Institute for Solid State Physics, the University of
Tokyo.
The simulations have been performed using an open-source implementation of the CTHYB solver~\cite{Hafermann2013} and some ALPS libraries~\cite{alps2}.
\end{acknowledgments}

\bibliography{dual}

\begin{thebibliography}{40}%
\makeatletter
\providecommand \@ifxundefined [1]{%
 \@ifx{#1\undefined}
}%
\providecommand \@ifnum [1]{%
 \ifnum #1\expandafter \@firstoftwo
 \else \expandafter \@secondoftwo
 \fi
}%
\providecommand \@ifx [1]{%
 \ifx #1\expandafter \@firstoftwo
 \else \expandafter \@secondoftwo
 \fi
}%
\providecommand \natexlab [1]{#1}%
\providecommand \enquote  [1]{``#1''}%
\providecommand \bibnamefont  [1]{#1}%
\providecommand \bibfnamefont [1]{#1}%
\providecommand \citenamefont [1]{#1}%
\providecommand \href@noop [0]{\@secondoftwo}%
\providecommand \href [0]{\begingroup \@sanitize@url \@href}%
\providecommand \@href[1]{\@@startlink{#1}\@@href}%
\providecommand \@@href[1]{\endgroup#1\@@endlink}%
\providecommand \@sanitize@url [0]{\catcode `\\12\catcode `\$12\catcode
  `\&12\catcode `\#12\catcode `\^12\catcode `\_12\catcode `\%12\relax}%
\providecommand \@@startlink[1]{}%
\providecommand \@@endlink[0]{}%
\providecommand \url  [0]{\begingroup\@sanitize@url \@url }%
\providecommand \@url [1]{\endgroup\@href {#1}{\urlprefix }}%
\providecommand \urlprefix  [0]{URL }%
\providecommand \Eprint [0]{\href }%
\providecommand \doibase [0]{http://dx.doi.org/}%
\providecommand \selectlanguage [0]{\@gobble}%
\providecommand \bibinfo  [0]{\@secondoftwo}%
\providecommand \bibfield  [0]{\@secondoftwo}%
\providecommand \translation [1]{[#1]}%
\providecommand \BibitemOpen [0]{}%
\providecommand \bibitemStop [0]{}%
\providecommand \bibitemNoStop [0]{.\EOS\space}%
\providecommand \EOS [0]{\spacefactor3000\relax}%
\providecommand \BibitemShut  [1]{\csname bibitem#1\endcsname}%
\let\auto@bib@innerbib\@empty
\bibitem [{\citenamefont {Bloch}\ \emph {et~al.}(2008)\citenamefont {Bloch},
  \citenamefont {Dalibard},\ and\ \citenamefont {Zwerger}}]{ultracold_gases}%
  \BibitemOpen
  \bibfield  {author} {\bibinfo {author} {\bibfnamefont {I.}~\bibnamefont
  {Bloch}}, \bibinfo {author} {\bibfnamefont {J.}~\bibnamefont {Dalibard}}, \
  and\ \bibinfo {author} {\bibfnamefont {W.}~\bibnamefont {Zwerger}},\ }\href
  {\doibase 10.1103/RevModPhys.80.885} {\bibfield  {journal} {\bibinfo
  {journal} {Rev. Mod. Phys.}\ }\textbf {\bibinfo {volume} {80}},\ \bibinfo
  {pages} {885} (\bibinfo {year} {2008})}\BibitemShut {NoStop}%
\bibitem [{\citenamefont {Izumi}\ \emph {et~al.}(2001)\citenamefont {Izumi},
  \citenamefont {Ogimoto}, \citenamefont {Konishi}, \citenamefont {Manako},
  \citenamefont {Kawasaki},\ and\ \citenamefont {Tokura}}]{IZUMI200153}%
  \BibitemOpen
  \bibfield  {author} {\bibinfo {author} {\bibfnamefont {M.}~\bibnamefont
  {Izumi}}, \bibinfo {author} {\bibfnamefont {Y.}~\bibnamefont {Ogimoto}},
  \bibinfo {author} {\bibfnamefont {Y.}~\bibnamefont {Konishi}}, \bibinfo
  {author} {\bibfnamefont {T.}~\bibnamefont {Manako}}, \bibinfo {author}
  {\bibfnamefont {M.}~\bibnamefont {Kawasaki}}, \ and\ \bibinfo {author}
  {\bibfnamefont {Y.}~\bibnamefont {Tokura}},\ }\href {\doibase
  http://dx.doi.org/10.1016/S0921-5107(01)00569-4} {\bibfield  {journal}
  {\bibinfo  {journal} {Materials Science and Engineering: B}\ }\textbf
  {\bibinfo {volume} {84}},\ \bibinfo {pages} {53 } (\bibinfo {year}
  {2001})}\BibitemShut {NoStop}%
\bibitem [{\citenamefont {Okamoto}\ and\ \citenamefont
  {Millis}(2004)}]{interface}%
  \BibitemOpen
  \bibfield  {author} {\bibinfo {author} {\bibfnamefont {S.}~\bibnamefont
  {Okamoto}}\ and\ \bibinfo {author} {\bibfnamefont {A.~J.}\ \bibnamefont
  {Millis}},\ }\href {\doibase 10.1038/nature02450} {\bibfield  {journal}
  {\bibinfo  {journal} {Nature}\ }\textbf {\bibinfo {volume} {428}},\ \bibinfo
  {pages} {630} (\bibinfo {year} {2004})}\BibitemShut {NoStop}%
\bibitem [{\citenamefont {K\"ohl}\ \emph {et~al.}(2005)\citenamefont {K\"ohl},
  \citenamefont {Moritz}, \citenamefont {St\"oferle}, \citenamefont
  {G\"unter},\ and\ \citenamefont {Esslinger}}]{inhomo_optical}%
  \BibitemOpen
  \bibfield  {author} {\bibinfo {author} {\bibfnamefont {M.}~\bibnamefont
  {K\"ohl}}, \bibinfo {author} {\bibfnamefont {H.}~\bibnamefont {Moritz}},
  \bibinfo {author} {\bibfnamefont {T.}~\bibnamefont {St\"oferle}}, \bibinfo
  {author} {\bibfnamefont {K.}~\bibnamefont {G\"unter}}, \ and\ \bibinfo
  {author} {\bibfnamefont {T.}~\bibnamefont {Esslinger}},\ }\href {\doibase
  10.1103/PhysRevLett.94.080403} {\bibfield  {journal} {\bibinfo  {journal}
  {Phys. Rev. Lett.}\ }\textbf {\bibinfo {volume} {94}},\ \bibinfo {pages}
  {080403} (\bibinfo {year} {2005})}\BibitemShut {NoStop}%
\bibitem [{\citenamefont {Ohtomo}\ \emph {et~al.}(2002)\citenamefont {Ohtomo},
  \citenamefont {Muller}, \citenamefont {Grazul},\ and\ \citenamefont
  {Hwang}}]{ohtomo2002artificial}%
  \BibitemOpen
  \bibfield  {author} {\bibinfo {author} {\bibfnamefont {A.}~\bibnamefont
  {Ohtomo}}, \bibinfo {author} {\bibfnamefont {D.}~\bibnamefont {Muller}},
  \bibinfo {author} {\bibfnamefont {J.}~\bibnamefont {Grazul}}, \ and\ \bibinfo
  {author} {\bibfnamefont {H.~Y.}\ \bibnamefont {Hwang}},\ }\href@noop {}
  {\bibfield  {journal} {\bibinfo  {journal} {Nature}\ }\textbf {\bibinfo
  {volume} {419}},\ \bibinfo {pages} {378} (\bibinfo {year}
  {2002})}\BibitemShut {NoStop}%
\bibitem [{\citenamefont {Ishimasa}\ \emph {et~al.}(2011)\citenamefont
  {Ishimasa}, \citenamefont {Tanaka},\ and\ \citenamefont
  {Kashimoto}}]{Ishimasa}%
  \BibitemOpen
  \bibfield  {author} {\bibinfo {author} {\bibfnamefont {T.}~\bibnamefont
  {Ishimasa}}, \bibinfo {author} {\bibfnamefont {Y.}~\bibnamefont {Tanaka}}, \
  and\ \bibinfo {author} {\bibfnamefont {S.}~\bibnamefont {Kashimoto}},\ }\href
  {\doibase 10.1080/14786435.2011.608732} {\bibfield  {journal} {\bibinfo
  {journal} {Philos. Mag.}\ }\textbf {\bibinfo {volume} {91}},\ \bibinfo
  {pages} {4218} (\bibinfo {year} {2011})}\BibitemShut {NoStop}%
\bibitem [{\citenamefont {Deguchi}\ \emph {et~al.}(2012)\citenamefont
  {Deguchi}, \citenamefont {Matsukawa}, \citenamefont {Sato}, \citenamefont
  {Hattori}, \citenamefont {Ishida}, \citenamefont {Takakura},\ and\
  \citenamefont {Ishimasa}}]{Deguchi12}%
  \BibitemOpen
  \bibfield  {author} {\bibinfo {author} {\bibfnamefont {K.}~\bibnamefont
  {Deguchi}}, \bibinfo {author} {\bibfnamefont {S.}~\bibnamefont {Matsukawa}},
  \bibinfo {author} {\bibfnamefont {N.~K.}\ \bibnamefont {Sato}}, \bibinfo
  {author} {\bibfnamefont {T.}~\bibnamefont {Hattori}}, \bibinfo {author}
  {\bibfnamefont {K.}~\bibnamefont {Ishida}}, \bibinfo {author} {\bibfnamefont
  {H.}~\bibnamefont {Takakura}}, \ and\ \bibinfo {author} {\bibfnamefont
  {T.}~\bibnamefont {Ishimasa}},\ }\href {\doibase 10.1038/nmat3432} {\bibfield
   {journal} {\bibinfo  {journal} {Nat. Mater.}\ }\textbf {\bibinfo {volume}
  {11}},\ \bibinfo {pages} {1013} (\bibinfo {year} {2012})}\BibitemShut
  {NoStop}%
\bibitem [{\citenamefont {Sarachik}\ \emph {et~al.}(1998)\citenamefont
  {Sarachik}, \citenamefont {Simonian}, \citenamefont {Kravchenko},
  \citenamefont {Bogdanovich}, \citenamefont {Dobrosavljevic},\ and\
  \citenamefont {Kotliar}}]{PhysRevB.58.6692}%
  \BibitemOpen
  \bibfield  {author} {\bibinfo {author} {\bibfnamefont {M.~P.}\ \bibnamefont
  {Sarachik}}, \bibinfo {author} {\bibfnamefont {D.}~\bibnamefont {Simonian}},
  \bibinfo {author} {\bibfnamefont {S.~V.}\ \bibnamefont {Kravchenko}},
  \bibinfo {author} {\bibfnamefont {S.}~\bibnamefont {Bogdanovich}}, \bibinfo
  {author} {\bibfnamefont {V.}~\bibnamefont {Dobrosavljevic}}, \ and\ \bibinfo
  {author} {\bibfnamefont {G.}~\bibnamefont {Kotliar}},\ }\href {\doibase
  10.1103/PhysRevB.58.6692} {\bibfield  {journal} {\bibinfo  {journal} {Phys.
  Rev. B}\ }\textbf {\bibinfo {volume} {58}},\ \bibinfo {pages} {6692}
  (\bibinfo {year} {1998})}\BibitemShut {NoStop}%
\bibitem [{\citenamefont {Seibold}\ \emph {et~al.}(1998)\citenamefont
  {Seibold}, \citenamefont {Sigmund},\ and\ \citenamefont
  {Hizhnyakov}}]{PhysRevB.57.6937}%
  \BibitemOpen
  \bibfield  {author} {\bibinfo {author} {\bibfnamefont {G.}~\bibnamefont
  {Seibold}}, \bibinfo {author} {\bibfnamefont {E.}~\bibnamefont {Sigmund}}, \
  and\ \bibinfo {author} {\bibfnamefont {V.}~\bibnamefont {Hizhnyakov}},\
  }\href {\doibase 10.1103/PhysRevB.57.6937} {\bibfield  {journal} {\bibinfo
  {journal} {Phys. Rev. B}\ }\textbf {\bibinfo {volume} {57}},\ \bibinfo
  {pages} {6937} (\bibinfo {year} {1998})}\BibitemShut {NoStop}%
\bibitem [{\citenamefont {Georges}\ \emph {et~al.}(1996)\citenamefont
  {Georges}, \citenamefont {Kotliar}, \citenamefont {Krauth},\ and\
  \citenamefont {Rozenberg}}]{Georges96}%
  \BibitemOpen
  \bibfield  {author} {\bibinfo {author} {\bibfnamefont {A.}~\bibnamefont
  {Georges}}, \bibinfo {author} {\bibfnamefont {G.}~\bibnamefont {Kotliar}},
  \bibinfo {author} {\bibfnamefont {W.}~\bibnamefont {Krauth}}, \ and\ \bibinfo
  {author} {\bibfnamefont {M.~J.}\ \bibnamefont {Rozenberg}},\ }\href {\doibase
  10.1103/RevModPhys.68.13} {\bibfield  {journal} {\bibinfo  {journal} {Rev.
  Mod. Phys.}\ }\textbf {\bibinfo {volume} {68}},\ \bibinfo {pages} {13}
  (\bibinfo {year} {1996})}\BibitemShut {NoStop}%
\bibitem [{\citenamefont {Potthoff}\ and\ \citenamefont
  {Nolting}(1999)}]{surface}%
  \BibitemOpen
  \bibfield  {author} {\bibinfo {author} {\bibfnamefont {M.}~\bibnamefont
  {Potthoff}}\ and\ \bibinfo {author} {\bibfnamefont {W.}~\bibnamefont
  {Nolting}},\ }\href {\doibase 10.1103/PhysRevB.59.2549} {\bibfield  {journal}
  {\bibinfo  {journal} {Phys. Rev. B}\ }\textbf {\bibinfo {volume} {59}},\
  \bibinfo {pages} {2549} (\bibinfo {year} {1999})}\BibitemShut {NoStop}%
\bibitem [{\citenamefont {Helmes}\ \emph {et~al.}(2008)\citenamefont {Helmes},
  \citenamefont {Costi},\ and\ \citenamefont {Rosch}}]{cold1}%
  \BibitemOpen
  \bibfield  {author} {\bibinfo {author} {\bibfnamefont {R.~W.}\ \bibnamefont
  {Helmes}}, \bibinfo {author} {\bibfnamefont {T.~A.}\ \bibnamefont {Costi}}, \
  and\ \bibinfo {author} {\bibfnamefont {A.}~\bibnamefont {Rosch}},\ }\href
  {\doibase 10.1103/PhysRevLett.100.056403} {\bibfield  {journal} {\bibinfo
  {journal} {Phys. Rev. Lett.}\ }\textbf {\bibinfo {volume} {100}},\ \bibinfo
  {pages} {056403} (\bibinfo {year} {2008})}\BibitemShut {NoStop}%
\bibitem [{\citenamefont {Snoek}\ \emph {et~al.}(2008)\citenamefont {Snoek},
  \citenamefont {Titvinidze}, \citenamefont {T{\H{o}}ke}, \citenamefont
  {Byczuk},\ and\ \citenamefont {Hofstetter}}]{cold2}%
  \BibitemOpen
  \bibfield  {author} {\bibinfo {author} {\bibfnamefont {M.}~\bibnamefont
  {Snoek}}, \bibinfo {author} {\bibfnamefont {I.}~\bibnamefont {Titvinidze}},
  \bibinfo {author} {\bibfnamefont {C.}~\bibnamefont {T{\H{o}}ke}}, \bibinfo
  {author} {\bibfnamefont {K.}~\bibnamefont {Byczuk}}, \ and\ \bibinfo {author}
  {\bibfnamefont {W.}~\bibnamefont {Hofstetter}},\ }\href
  {http://stacks.iop.org/1367-2630/10/i=9/a=093008} {\bibfield  {journal}
  {\bibinfo  {journal} {New J. Phys.}\ }\textbf {\bibinfo {volume} {10}},\
  \bibinfo {pages} {093008} (\bibinfo {year} {2008})}\BibitemShut {NoStop}%
\bibitem [{\citenamefont {Koga}\ \emph {et~al.}(2008)\citenamefont {Koga},
  \citenamefont {Higashiyama}, \citenamefont {Inaba}, \citenamefont {Suga},\
  and\ \citenamefont {Kawakami}}]{cold3}%
  \BibitemOpen
  \bibfield  {author} {\bibinfo {author} {\bibfnamefont {A.}~\bibnamefont
  {Koga}}, \bibinfo {author} {\bibfnamefont {T.}~\bibnamefont {Higashiyama}},
  \bibinfo {author} {\bibfnamefont {K.}~\bibnamefont {Inaba}}, \bibinfo
  {author} {\bibfnamefont {S.}~\bibnamefont {Suga}}, \ and\ \bibinfo {author}
  {\bibfnamefont {N.}~\bibnamefont {Kawakami}},\ }\href {\doibase
  10.1143/JPSJ.77.073602} {\bibfield  {journal} {\bibinfo  {journal} {J. Phys.
  Soc. Jpn.}\ }\textbf {\bibinfo {volume} {77}},\ \bibinfo {pages} {073602}
  (\bibinfo {year} {2008})}\BibitemShut {NoStop}%
\bibitem [{\citenamefont {Takemori}\ and\ \citenamefont
  {Koga}(2015{\natexlab{a}})}]{Takemori}%
  \BibitemOpen
  \bibfield  {author} {\bibinfo {author} {\bibfnamefont {N.}~\bibnamefont
  {Takemori}}\ and\ \bibinfo {author} {\bibfnamefont {A.}~\bibnamefont
  {Koga}},\ }\href {\doibase 10.7566/JPSJ.84.023701} {\bibfield  {journal}
  {\bibinfo  {journal} {J. Phys. Soc. Jpn.}\ }\textbf {\bibinfo {volume}
  {84}},\ \bibinfo {pages} {023701} (\bibinfo {year}
  {2015}{\natexlab{a}})}\BibitemShut {NoStop}%
\bibitem [{\citenamefont {Takemura}\ \emph {et~al.}(2015)\citenamefont
  {Takemura}, \citenamefont {Takemori},\ and\ \citenamefont {Koga}}]{Takemura}%
  \BibitemOpen
  \bibfield  {author} {\bibinfo {author} {\bibfnamefont {S.}~\bibnamefont
  {Takemura}}, \bibinfo {author} {\bibfnamefont {N.}~\bibnamefont {Takemori}},
  \ and\ \bibinfo {author} {\bibfnamefont {A.}~\bibnamefont {Koga}},\ }\href
  {\doibase 10.1103/PhysRevB.91.165114} {\bibfield  {journal} {\bibinfo
  {journal} {Phys. Rev. B}\ }\textbf {\bibinfo {volume} {91}},\ \bibinfo
  {pages} {165114} (\bibinfo {year} {2015})}\BibitemShut {NoStop}%
\bibitem [{\citenamefont {{Rohringer}}\ \emph {et~al.}(2017)\citenamefont
  {{Rohringer}}, \citenamefont {{Hafermann}}, \citenamefont {{Toschi}},
  \citenamefont {{Katanin}}, \citenamefont {{Antipov}}, \citenamefont
  {{Katsnelson}}, \citenamefont {{Lichtenstein}}, \citenamefont {{Rubtsov}},\
  and\ \citenamefont {{Held}}}]{Rohringer17}%
  \BibitemOpen
  \bibfield  {author} {\bibinfo {author} {\bibfnamefont {G.}~\bibnamefont
  {{Rohringer}}}, \bibinfo {author} {\bibfnamefont {H.}~\bibnamefont
  {{Hafermann}}}, \bibinfo {author} {\bibfnamefont {A.}~\bibnamefont
  {{Toschi}}}, \bibinfo {author} {\bibfnamefont {A.~A.}\ \bibnamefont
  {{Katanin}}}, \bibinfo {author} {\bibfnamefont {A.~E.}\ \bibnamefont
  {{Antipov}}}, \bibinfo {author} {\bibfnamefont {M.~I.}\ \bibnamefont
  {{Katsnelson}}}, \bibinfo {author} {\bibfnamefont {A.~I.}\ \bibnamefont
  {{Lichtenstein}}}, \bibinfo {author} {\bibfnamefont {A.~N.}\ \bibnamefont
  {{Rubtsov}}}, \ and\ \bibinfo {author} {\bibfnamefont {K.}~\bibnamefont
  {{Held}}},\ }\href@noop {} {\bibfield  {journal} {\bibinfo  {journal} {ArXiv
  e-prints}\ } (\bibinfo {year} {2017})},\ \Eprint
  {http://arxiv.org/abs/1705.00024} {arXiv:1705.00024 [cond-mat.str-el]}
  \BibitemShut {NoStop}%
\bibitem [{\citenamefont {Kusunose}(2006)}]{Kusunose}%
  \BibitemOpen
  \bibfield  {author} {\bibinfo {author} {\bibfnamefont {H.}~\bibnamefont
  {Kusunose}},\ }\href {\doibase 10.1143/JPSJ.75.054713} {\bibfield  {journal}
  {\bibinfo  {journal} {J. Phys. Soc. Jpn.}\ }\textbf {\bibinfo {volume}
  {75}},\ \bibinfo {pages} {054713} (\bibinfo {year} {2006})}\BibitemShut
  {NoStop}%
\bibitem [{\citenamefont {Valli}\ \emph {et~al.}(2010)\citenamefont {Valli},
  \citenamefont {Sangiovanni}, \citenamefont {Gunnarsson}, \citenamefont
  {Toschi},\ and\ \citenamefont {Held}}]{Valli2010}%
  \BibitemOpen
  \bibfield  {author} {\bibinfo {author} {\bibfnamefont {A.}~\bibnamefont
  {Valli}}, \bibinfo {author} {\bibfnamefont {G.}~\bibnamefont {Sangiovanni}},
  \bibinfo {author} {\bibfnamefont {O.}~\bibnamefont {Gunnarsson}}, \bibinfo
  {author} {\bibfnamefont {A.}~\bibnamefont {Toschi}}, \ and\ \bibinfo {author}
  {\bibfnamefont {K.}~\bibnamefont {Held}},\ }\href {\doibase
  10.1103/PhysRevLett.104.246402} {\bibfield  {journal} {\bibinfo  {journal}
  {Phys. Rev. Lett.}\ }\textbf {\bibinfo {volume} {104}},\ \bibinfo {pages}
  {246402} (\bibinfo {year} {2010})}\BibitemShut {NoStop}%
\bibitem [{\citenamefont {Valli}\ \emph {et~al.}(2012)\citenamefont {Valli},
  \citenamefont {Sangiovanni}, \citenamefont {Toschi},\ and\ \citenamefont
  {Held}}]{Valli2012}%
  \BibitemOpen
  \bibfield  {author} {\bibinfo {author} {\bibfnamefont {A.}~\bibnamefont
  {Valli}}, \bibinfo {author} {\bibfnamefont {G.}~\bibnamefont {Sangiovanni}},
  \bibinfo {author} {\bibfnamefont {A.}~\bibnamefont {Toschi}}, \ and\ \bibinfo
  {author} {\bibfnamefont {K.}~\bibnamefont {Held}},\ }\href {\doibase
  10.1103/PhysRevB.86.115418} {\bibfield  {journal} {\bibinfo  {journal} {Phys.
  Rev. B}\ }\textbf {\bibinfo {volume} {86}},\ \bibinfo {pages} {115418}
  (\bibinfo {year} {2012})}\BibitemShut {NoStop}%
\bibitem [{\citenamefont {Takemori}\ \emph {et~al.}(2016)\citenamefont
  {Takemori}, \citenamefont {Koga},\ and\ \citenamefont
  {Hafermann}}]{Takemoridualproc}%
  \BibitemOpen
  \bibfield  {author} {\bibinfo {author} {\bibfnamefont {N.}~\bibnamefont
  {Takemori}}, \bibinfo {author} {\bibfnamefont {A.}~\bibnamefont {Koga}}, \
  and\ \bibinfo {author} {\bibfnamefont {H.}~\bibnamefont {Hafermann}},\ }\href
  {http://stacks.iop.org/1742-6596/683/i=1/a=012040} {\bibfield  {journal}
  {\bibinfo  {journal} {J. Phys.: Conf. Ser.}\ }\textbf {\bibinfo {volume}
  {683}},\ \bibinfo {pages} {012040} (\bibinfo {year} {2016})}\BibitemShut
  {NoStop}%
\bibitem [{\citenamefont {Rubtsov}\ \emph {et~al.}(2008)\citenamefont
  {Rubtsov}, \citenamefont {Katsnelson},\ and\ \citenamefont
  {Lichtenstein}}]{Rubtsov}%
  \BibitemOpen
  \bibfield  {author} {\bibinfo {author} {\bibfnamefont {A.~N.}\ \bibnamefont
  {Rubtsov}}, \bibinfo {author} {\bibfnamefont {M.~I.}\ \bibnamefont
  {Katsnelson}}, \ and\ \bibinfo {author} {\bibfnamefont {A.~I.}\ \bibnamefont
  {Lichtenstein}},\ }\href {\doibase 10.1103/PhysRevB.77.033101} {\bibfield
  {journal} {\bibinfo  {journal} {Phys. Rev. B}\ }\textbf {\bibinfo {volume}
  {77}},\ \bibinfo {pages} {033101} (\bibinfo {year} {2008})}\BibitemShut
  {NoStop}%
\bibitem [{\citenamefont {Otsuki}\ \emph {et~al.}(2014)\citenamefont {Otsuki},
  \citenamefont {Hafermann},\ and\ \citenamefont {Lichtenstein}}]{Otsuki}%
  \BibitemOpen
  \bibfield  {author} {\bibinfo {author} {\bibfnamefont {J.}~\bibnamefont
  {Otsuki}}, \bibinfo {author} {\bibfnamefont {H.}~\bibnamefont {Hafermann}}, \
  and\ \bibinfo {author} {\bibfnamefont {A.~I.}\ \bibnamefont {Lichtenstein}},\
  }\href {\doibase 10.1103/PhysRevB.90.235132} {\bibfield  {journal} {\bibinfo
  {journal} {Phys. Rev. B}\ }\textbf {\bibinfo {volume} {90}},\ \bibinfo
  {pages} {235132} (\bibinfo {year} {2014})}\BibitemShut {NoStop}%
\bibitem [{\citenamefont {Hafermann}(2010)}]{Hartmut}%
  \BibitemOpen
  \bibfield  {author} {\bibinfo {author} {\bibfnamefont {H.}~\bibnamefont
  {Hafermann}},\ }\href@noop {} {\emph {\bibinfo {title} {Numerical Approaches
  to Spatial Correlations in Strongly Interacting Fermion Systems}}}\ (\bibinfo
   {publisher} {Cuvillier Verlag, G\"ottingen},\ \bibinfo {year}
  {2010})\BibitemShut {NoStop}%
\bibitem [{\citenamefont {Li}\ \emph {et~al.}(2014)\citenamefont {Li},
  \citenamefont {Antipov}, \citenamefont {Rubtsov}, \citenamefont {Kirchner},\
  and\ \citenamefont {Hanke}}]{Li2014}%
  \BibitemOpen
  \bibfield  {author} {\bibinfo {author} {\bibfnamefont {G.}~\bibnamefont
  {Li}}, \bibinfo {author} {\bibfnamefont {A.~E.}\ \bibnamefont {Antipov}},
  \bibinfo {author} {\bibfnamefont {A.~N.}\ \bibnamefont {Rubtsov}}, \bibinfo
  {author} {\bibfnamefont {S.}~\bibnamefont {Kirchner}}, \ and\ \bibinfo
  {author} {\bibfnamefont {W.}~\bibnamefont {Hanke}},\ }\href {\doibase
  10.1103/PhysRevB.89.161118} {\bibfield  {journal} {\bibinfo  {journal} {Phys.
  Rev. B}\ }\textbf {\bibinfo {volume} {89}},\ \bibinfo {pages} {161118}
  (\bibinfo {year} {2014})}\BibitemShut {NoStop}%
\bibitem [{\citenamefont {Hafermann}\ \emph {et~al.}(2009)\citenamefont
  {Hafermann}, \citenamefont {Li}, \citenamefont {Rubtsov}, \citenamefont
  {Katsnelson}, \citenamefont {Lichtenstein},\ and\ \citenamefont
  {Monien}}]{Hafermann09}%
  \BibitemOpen
  \bibfield  {author} {\bibinfo {author} {\bibfnamefont {H.}~\bibnamefont
  {Hafermann}}, \bibinfo {author} {\bibfnamefont {G.}~\bibnamefont {Li}},
  \bibinfo {author} {\bibfnamefont {A.~N.}\ \bibnamefont {Rubtsov}}, \bibinfo
  {author} {\bibfnamefont {M.~I.}\ \bibnamefont {Katsnelson}}, \bibinfo
  {author} {\bibfnamefont {A.~I.}\ \bibnamefont {Lichtenstein}}, \ and\
  \bibinfo {author} {\bibfnamefont {H.}~\bibnamefont {Monien}},\ }\href
  {\doibase 10.1103/PhysRevLett.102.206401} {\bibfield  {journal} {\bibinfo
  {journal} {Phys. Rev. Lett.}\ }\textbf {\bibinfo {volume} {102}},\ \bibinfo
  {pages} {206401} (\bibinfo {year} {2009})}\BibitemShut {NoStop}%
\bibitem [{\citenamefont {Antipov}\ \emph {et~al.}(2014)\citenamefont
  {Antipov}, \citenamefont {Gull},\ and\ \citenamefont
  {Kirchner}}]{FK_critical_exponent}%
  \BibitemOpen
  \bibfield  {author} {\bibinfo {author} {\bibfnamefont {A.~E.}\ \bibnamefont
  {Antipov}}, \bibinfo {author} {\bibfnamefont {E.}~\bibnamefont {Gull}}, \
  and\ \bibinfo {author} {\bibfnamefont {S.}~\bibnamefont {Kirchner}},\ }\href
  {\doibase 10.1103/PhysRevLett.112.226401} {\bibfield  {journal} {\bibinfo
  {journal} {Phys. Rev. Lett.}\ }\textbf {\bibinfo {volume} {112}},\ \bibinfo
  {pages} {226401} (\bibinfo {year} {2014})}\BibitemShut {NoStop}%
\bibitem [{\citenamefont {Hirschmeier}\ \emph {et~al.}(2015)\citenamefont
  {Hirschmeier}, \citenamefont {Hafermann}, \citenamefont {Gull}, \citenamefont
  {Lichtenstein},\ and\ \citenamefont {Antipov}}]{Hirschmeier2015}%
  \BibitemOpen
  \bibfield  {author} {\bibinfo {author} {\bibfnamefont {D.}~\bibnamefont
  {Hirschmeier}}, \bibinfo {author} {\bibfnamefont {H.}~\bibnamefont
  {Hafermann}}, \bibinfo {author} {\bibfnamefont {E.}~\bibnamefont {Gull}},
  \bibinfo {author} {\bibfnamefont {A.~I.}\ \bibnamefont {Lichtenstein}}, \
  and\ \bibinfo {author} {\bibfnamefont {A.~E.}\ \bibnamefont {Antipov}},\
  }\href {\doibase 10.1103/PhysRevB.92.144409} {\bibfield  {journal} {\bibinfo
  {journal} {Phys. Rev. B}\ }\textbf {\bibinfo {volume} {92}},\ \bibinfo
  {pages} {144409} (\bibinfo {year} {2015})}\BibitemShut {NoStop}%
\bibitem [{\citenamefont {Hafermann}\ \emph {et~al.}(2013)\citenamefont
  {Hafermann}, \citenamefont {Werner},\ and\ \citenamefont
  {Gull}}]{Hafermann2013}%
  \BibitemOpen
  \bibfield  {author} {\bibinfo {author} {\bibfnamefont {H.}~\bibnamefont
  {Hafermann}}, \bibinfo {author} {\bibfnamefont {P.}~\bibnamefont {Werner}}, \
  and\ \bibinfo {author} {\bibfnamefont {E.}~\bibnamefont {Gull}},\ }\href
  {\doibase 10.1016/j.cpc.2012.12.013} {\bibfield  {journal} {\bibinfo
  {journal} {Computer Physics Communications}\ }\textbf {\bibinfo {volume}
  {184}},\ \bibinfo {pages} {1280 } (\bibinfo {year} {2013})}\BibitemShut
  {NoStop}%
\bibitem [{\citenamefont {Rozenberg}\ \emph {et~al.}(1999)\citenamefont
  {Rozenberg}, \citenamefont {Chitra},\ and\ \citenamefont {Kotliar}}]{single}%
  \BibitemOpen
  \bibfield  {author} {\bibinfo {author} {\bibfnamefont {M.~J.}\ \bibnamefont
  {Rozenberg}}, \bibinfo {author} {\bibfnamefont {R.}~\bibnamefont {Chitra}}, \
  and\ \bibinfo {author} {\bibfnamefont {G.}~\bibnamefont {Kotliar}},\ }\href
  {\doibase 10.1103/PhysRevLett.83.3498} {\bibfield  {journal} {\bibinfo
  {journal} {Phys. Rev. Lett.}\ }\textbf {\bibinfo {volume} {83}},\ \bibinfo
  {pages} {3498} (\bibinfo {year} {1999})}\BibitemShut {NoStop}%
\bibitem [{\citenamefont {Takemori}\ and\ \citenamefont
  {Koga}(2015{\natexlab{b}})}]{Takemoriproc}%
  \BibitemOpen
  \bibfield  {author} {\bibinfo {author} {\bibfnamefont {N.}~\bibnamefont
  {Takemori}}\ and\ \bibinfo {author} {\bibfnamefont {A.}~\bibnamefont
  {Koga}},\ }\href {http://stacks.iop.org/1742-6596/592/i=1/a=012038}
  {\bibfield  {journal} {\bibinfo  {journal} {J. Phys.: Conf. Ser.}\ }\textbf
  {\bibinfo {volume} {592}},\ \bibinfo {pages} {012038} (\bibinfo {year}
  {2015}{\natexlab{b}})}\BibitemShut {NoStop}%
\bibitem [{\citenamefont {Werner}()}]{Werner_private}%
  \BibitemOpen
  \bibfield  {author} {\bibinfo {author} {\bibfnamefont {P.}~\bibnamefont
  {Werner}},\ }\href@noop {} {}\bibinfo {howpublished} {private
  communication}\BibitemShut {NoStop}%
\bibitem [{\citenamefont {Park}\ \emph {et~al.}(2008)\citenamefont {Park},
  \citenamefont {Haule},\ and\ \citenamefont {Kotliar}}]{square_CDMFT}%
  \BibitemOpen
  \bibfield  {author} {\bibinfo {author} {\bibfnamefont {H.}~\bibnamefont
  {Park}}, \bibinfo {author} {\bibfnamefont {K.}~\bibnamefont {Haule}}, \ and\
  \bibinfo {author} {\bibfnamefont {G.}~\bibnamefont {Kotliar}},\ }\href
  {\doibase 10.1103/PhysRevLett.101.186403} {\bibfield  {journal} {\bibinfo
  {journal} {Phys. Rev. Lett.}\ }\textbf {\bibinfo {volume} {101}},\ \bibinfo
  {pages} {186403} (\bibinfo {year} {2008})}\BibitemShut {NoStop}%
\bibitem [{\citenamefont {Veki\ifmmode~\acute{c}\else \'{c}\fi{}}\ and\
  \citenamefont {White}(1993)}]{squareQMC}%
  \BibitemOpen
  \bibfield  {author} {\bibinfo {author} {\bibfnamefont {M.}~\bibnamefont
  {Veki\ifmmode~\acute{c}\else \'{c}\fi{}}}\ and\ \bibinfo {author}
  {\bibfnamefont {S.~R.}\ \bibnamefont {White}},\ }\href {\doibase
  10.1103/PhysRevB.47.1160} {\bibfield  {journal} {\bibinfo  {journal} {Phys.
  Rev. B}\ }\textbf {\bibinfo {volume} {47}},\ \bibinfo {pages} {1160}
  (\bibinfo {year} {1993})}\BibitemShut {NoStop}%
\bibitem [{\citenamefont {Sakai}\ \emph {et~al.}(2017)\citenamefont {Sakai},
  \citenamefont {Takemori}, \citenamefont {Koga},\ and\ \citenamefont
  {Arita}}]{Sakai}%
  \BibitemOpen
  \bibfield  {author} {\bibinfo {author} {\bibfnamefont {S.}~\bibnamefont
  {Sakai}}, \bibinfo {author} {\bibfnamefont {N.}~\bibnamefont {Takemori}},
  \bibinfo {author} {\bibfnamefont {A.}~\bibnamefont {Koga}}, \ and\ \bibinfo
  {author} {\bibfnamefont {R.}~\bibnamefont {Arita}},\ }\href {\doibase
  10.1103/PhysRevB.95.024509} {\bibfield  {journal} {\bibinfo  {journal} {Phys.
  Rev. B}\ }\textbf {\bibinfo {volume} {95}},\ \bibinfo {pages} {024509}
  (\bibinfo {year} {2017})}\BibitemShut {NoStop}%
\bibitem [{\citenamefont {Koga}\ and\ \citenamefont
  {Tsunetsugu}(2017)}]{koga-tsunetsugu}%
  \BibitemOpen
  \bibfield  {author} {\bibinfo {author} {\bibfnamefont {A.}~\bibnamefont
  {Koga}}\ and\ \bibinfo {author} {\bibfnamefont {H.}~\bibnamefont
  {Tsunetsugu}},\ }\href {\doibase 10.1103/PhysRevB.96.214402} {\bibfield
  {journal} {\bibinfo  {journal} {Phys. Rev. B}\ }\textbf {\bibinfo {volume}
  {96}},\ \bibinfo {pages} {214402} (\bibinfo {year} {2017})}\BibitemShut
  {NoStop}%
\bibitem [{\citenamefont {De~Bruijn}(1981{\natexlab{a}})}]{Bruijn1}%
  \BibitemOpen
  \bibfield  {author} {\bibinfo {author} {\bibfnamefont {N.}~\bibnamefont
  {De~Bruijn}},\ }\href@noop {} {\bibfield  {journal} {\bibinfo  {journal}
  {Indag. Math.}\ }\textbf {\bibinfo {volume} {84}},\ \bibinfo {pages} {39}
  (\bibinfo {year} {1981}{\natexlab{a}})}\BibitemShut {NoStop}%
\bibitem [{\citenamefont {De~Bruijn}(1981{\natexlab{b}})}]{Bruijn2}%
  \BibitemOpen
  \bibfield  {author} {\bibinfo {author} {\bibfnamefont {N.}~\bibnamefont
  {De~Bruijn}},\ }\href@noop {} {\bibfield  {journal} {\bibinfo  {journal}
  {Indag. Math.}\ }\textbf {\bibinfo {volume} {84}},\ \bibinfo {pages} {53}
  (\bibinfo {year} {1981}{\natexlab{b}})}\BibitemShut {NoStop}%
\bibitem [{\citenamefont {Shinaoka}\ \emph {et~al.}(2017)\citenamefont
  {Shinaoka}, \citenamefont {Otsuki}, \citenamefont {Ohzeki},\ and\
  \citenamefont {Yoshimi}}]{shinaoka2017compression}%
  \BibitemOpen
  \bibfield  {author} {\bibinfo {author} {\bibfnamefont {H.}~\bibnamefont
  {Shinaoka}}, \bibinfo {author} {\bibfnamefont {J.}~\bibnamefont {Otsuki}},
  \bibinfo {author} {\bibfnamefont {M.}~\bibnamefont {Ohzeki}}, \ and\ \bibinfo
  {author} {\bibfnamefont {K.}~\bibnamefont {Yoshimi}},\ }\href {\doibase
  10.1103/PhysRevB.96.035147} {\bibfield  {journal} {\bibinfo  {journal} {Phys.
  Rev. B}\ }\textbf {\bibinfo {volume} {96}},\ \bibinfo {pages} {035147}
  (\bibinfo {year} {2017})}\BibitemShut {NoStop}%
\bibitem [{\citenamefont {Bauer}\ \emph {et~al.}(2011)\citenamefont {Bauer},
  \citenamefont {Carr}, \citenamefont {Evertz}, \citenamefont {Feiguin},
  \citenamefont {Freire}, \citenamefont {Fuchs}, \citenamefont {Gamper},
  \citenamefont {Gukelberger}, \citenamefont {Gull}, \citenamefont {Guertler},
  \citenamefont {Hehn}, \citenamefont {Igarashi}, \citenamefont {Isakov},
  \citenamefont {Koop}, \citenamefont {Ma}, \citenamefont {Mates},
  \citenamefont {Matsuo}, \citenamefont {Parcollet}, \citenamefont
  {Paw{\l}owski}, \citenamefont {Picon}, \citenamefont {Pollet}, \citenamefont
  {Santos}, \citenamefont {Scarola}, \citenamefont {Schollw{\"o}ck},
  \citenamefont {Silva}, \citenamefont {Surer}, \citenamefont {Todo},
  \citenamefont {Trebst}, \citenamefont {Troyer}, \citenamefont {Wall},
  \citenamefont {Werner},\ and\ \citenamefont {Wessel}}]{alps2}%
  \BibitemOpen
  \bibfield  {author} {\bibinfo {author} {\bibfnamefont {B.}~\bibnamefont
  {Bauer}}, \bibinfo {author} {\bibfnamefont {L.~D.}\ \bibnamefont {Carr}},
  \bibinfo {author} {\bibfnamefont {H.~G.}\ \bibnamefont {Evertz}}, \bibinfo
  {author} {\bibfnamefont {A.}~\bibnamefont {Feiguin}}, \bibinfo {author}
  {\bibfnamefont {J.}~\bibnamefont {Freire}}, \bibinfo {author} {\bibfnamefont
  {S.}~\bibnamefont {Fuchs}}, \bibinfo {author} {\bibfnamefont
  {L.}~\bibnamefont {Gamper}}, \bibinfo {author} {\bibfnamefont
  {J.}~\bibnamefont {Gukelberger}}, \bibinfo {author} {\bibfnamefont
  {E.}~\bibnamefont {Gull}}, \bibinfo {author} {\bibfnamefont {S.}~\bibnamefont
  {Guertler}}, \bibinfo {author} {\bibfnamefont {A.}~\bibnamefont {Hehn}},
  \bibinfo {author} {\bibfnamefont {R.}~\bibnamefont {Igarashi}}, \bibinfo
  {author} {\bibfnamefont {S.~V.}\ \bibnamefont {Isakov}}, \bibinfo {author}
  {\bibfnamefont {D.}~\bibnamefont {Koop}}, \bibinfo {author} {\bibfnamefont
  {P.~N.}\ \bibnamefont {Ma}}, \bibinfo {author} {\bibfnamefont
  {P.}~\bibnamefont {Mates}}, \bibinfo {author} {\bibfnamefont
  {H.}~\bibnamefont {Matsuo}}, \bibinfo {author} {\bibfnamefont
  {O.}~\bibnamefont {Parcollet}}, \bibinfo {author} {\bibfnamefont
  {G.}~\bibnamefont {Paw{\l}owski}}, \bibinfo {author} {\bibfnamefont {J.~D.}\
  \bibnamefont {Picon}}, \bibinfo {author} {\bibfnamefont {L.}~\bibnamefont
  {Pollet}}, \bibinfo {author} {\bibfnamefont {E.}~\bibnamefont {Santos}},
  \bibinfo {author} {\bibfnamefont {V.~W.}\ \bibnamefont {Scarola}}, \bibinfo
  {author} {\bibfnamefont {U.}~\bibnamefont {Schollw{\"o}ck}}, \bibinfo
  {author} {\bibfnamefont {C.}~\bibnamefont {Silva}}, \bibinfo {author}
  {\bibfnamefont {B.}~\bibnamefont {Surer}}, \bibinfo {author} {\bibfnamefont
  {S.}~\bibnamefont {Todo}}, \bibinfo {author} {\bibfnamefont {S.}~\bibnamefont
  {Trebst}}, \bibinfo {author} {\bibfnamefont {M.}~\bibnamefont {Troyer}},
  \bibinfo {author} {\bibfnamefont {M.~L.}\ \bibnamefont {Wall}}, \bibinfo
  {author} {\bibfnamefont {P.}~\bibnamefont {Werner}}, \ and\ \bibinfo {author}
  {\bibfnamefont {S.}~\bibnamefont {Wessel}},\ }\href
  {http://stacks.iop.org/1742-5468/2011/i=05/a=P05001} {\bibfield  {journal}
  {\bibinfo  {journal} {J. Stat. Mech.: Theory E}\ }\textbf {\bibinfo {volume}
  {2011}},\ \bibinfo {pages} {P05001} (\bibinfo {year} {2011})}\BibitemShut
  {NoStop}%
\end{thebibliography}%
\end{document}